# Hardware Architecture for Single Iteration Reconstruction Algorithm

Andjela Draganić, Irena Orović, Nedjeljko Lekić, Miloš Daković, Srdjan Stanković
University of Montenegro
Faculty of Electrical Engineering
Podgorica, Montenegro
e-mail: andjelad@ac.me

*Abstract* - **A hardware architecture for the single iteration algorithm is proposed in this paper. Single iteration algorithm enables reconstruction of the full signal when small number of signal samples is available. The algorithm is based on the threshold calculation, and allows distinguishing between signal components and noise that appears as a consequence of missing samples. The proposed system for hardware realization is divided into three parts, each part with different functionality. The system is suitable for the FPGA realization. Realization of the blocks for which there are no standard components in FPGA, is discussed as well.**

I. INTRODUCTION

The reconstruction of missing samples has attracted great attention of the researches in the recent years [1]-[3]. Missing samples in the signal can occur by omitting samples during the acquisition process [4]. In the noisy signal cases, we can consider corrupted samples as missing ones if we are able to detect their positions [5],[6]. If the intentionally omitted samples or noisy samples are randomly distributed in the signal, then Compressive Sensing (CS) approach can be used for full signal reconstruction.

CS provides successful reconstruction using small set of available signal samples. Except random distribution of the available samples, CS requires the sparsity condition to be satisfied. It is related to the signal nature, i.e., there exists certain domain where signal has most of its coefficients equal to zero. In the other words, signal energy is concentrated within small number of coefficients in certain basis $\Psi$. Therefore, the area of CS application is large [4]. Vector of available measurements $v$ is called measurement vector, and can be formed as follows:

$$v_{M \times 1} = \Psi_{N \times N} \Theta_{M \times N} x_{N \times 1}, \qquad (1)$$

where $x$ is a signal, $M$ denotes number of available measurements, while $N$ is signal length. Matrix $\Theta$ is used to model random selections of the original signal samples. Smaller number of available samples compared to signal length in (1), causes undetermined system of equation which has to be solved in order to recover signal. Therefore, in order to obtain unique solution of the undetermined system, optimization algorithms are used [7]-[12].

Our focus in this paper will be on single iteration reconstruction (SIRA) algorithm for reconstruction of under-sampled signals [12]. The algorithm calculates threshold that allows distinguishing between noisy and signal components. The threshold calculation is based on predefined probability of error and determines minimal number of samples required for the signal reconstruction, in one iteration. In real signal processing applications, the hardware implementations are required to provide real-time solutions [13]-[20]. Hardware architecture for the implementation of single iteration CS algorithm is presented in the paper. Architecture is suitable for the realization in the FPGA. Moreover, the proposed architecture can be used to improve the performance of existing hardware, [17], by including CS based modules for signal reconstruction.

The paper is structured as follows: The second section describes the single iteration algorithm. The proposed hardware block scheme is described in section III while the detailed block descriptions with figures can be found in section IV. Concluding remarks are given in the section V.

II. ALGORITHM FOR THE SIGNAL RECONSTRUCTION

The single iteration algorithm for the CS signal reconstruction and block scheme for the FPGA implementation is considered in this paper. The algorithm is based on threshold calculation and choice of initial discrete Fourier transform (DFT) components that are above defined threshold. Initial DFT is calculated based on set of available signal samples. It is shown that components above the threshold correspond to the signal components, while the components below the threshold are considered as noise. Let us summarize the algorithm steps. Assume signal in the form:

$$x(n) = \sum_{i=1}^{K} A_i e^{j(2\pi k_i n / N)}, \qquad (2)$$

where $K$ denotes number of signal components, $N$ is signal length, $A_i$ and $k_i$ are amplitudes and frequencies of signal components. Assume further that some of the signal components at randomly selected positions are missing, or they are intentionally omitted. As a consequence of the

This work is supported by the Montenegrin Ministry of Science, project grant funded by the World Bank loan: CS-ICT "New ICT Compressive sensing based trends applied to: multimedia, biomedicine and communications".

missing samples, the noise appears in the signal. It is shown that the variance of this noise can be modeled as [10]:

$$\text{var} = M \frac{N_a}{N-1}(A_1^2 + A_2^2 + ... + A_K^2), \quad (3)$$

where $M$ is number of missing samples, while $N_a$ is number of available samples. This variance will be used in threshold calculation:

$$T = \frac{1}{N}\left(-\text{var}^2 \log_{10}(1 - \sqrt[N]{P})\right)^{1/2}, \quad (4)$$

The $P$ is the probability that ($N$-$K$) components that correspond to noise, are lower than the threshold. Note that in the above relation the approximation is used: $N$-th root instead of ($N$-$K$)-th root is used, since $K<<N$ [10]. The samples positions in the initial DFT that are above the threshold are used for the calculation of the exact DFT signal amplitudes, while the other frequency positions are filled with zeros. Vector of initial DFT is formed using the available time-domain signal samples, i.e. using the vector of measurements. Let $v_{M\times 1}$ denotes measurement vector. The initial DFT vector $V$ is then formed as:

$$V(f) = \sum_{a=1}^{N_a} v(a)e^{-j2\pi fa/N}, f = 1,...,N. \quad (5)$$

Positions of the components above the threshold are obtained by using following relation:

$$pos = \arg\{|V| > T\}. \quad (6)$$

In this way, only the frequency positions of the signal components are founded. To obtain the exact amplitudes of the components on the corresponding frequencies, the minimization problem has to be solved.

Let us form the CS matrix, used in an optimization problem. By subsampling DFT matrix $A_{N\times N}$, CS DFT matrix is obtained. Sub-sampling is done by rows and by columns. Row subsampling implies selection of the rows that correspond to the $M$ available measurements. Column subsampling selects columns that correspond to the positions of the samples above the threshold. The optimization problem to be solved is then formed as follows:

$$X = (A_{CS}^* A_{CS})^{-1}(A_{CS}^* v), \quad (7)$$

where CS matrix $A_{CS}$ is formed as $A(M, pos)$ and $A^*_{CS}$ is Hermitian transpose of the matrix $A_{CS}$.

### III. COMMENTS ON THE ARCHITECTURE SUITABLE FOR THE FPGA IMPLEMENTATION

In this part, the main building blocks required for the hardware implementation of the algorithm will be commented. The scheme can be divided into three parts:

**Part 1:** forming measurement vector and finding the positions above the threshold in the initial DFT vector; **Part 2:** forming the matrices used in optimization problem and **Part 3:** optimization problem solving.

Part 1 requires block for the randomization and selection part of the randomly permuted signal samples, block for DFT calculation, block for the threshold calculation and comparator. Note that the full signal is fed to the random generator. In real cases, only part of the signal samples is available. Therefore, case when we have all signal samples available can be used for signal denoising, when signal is corrupted with Gaussian noise. Threshold calculation requires logarithmic and power function calculation, and will be explained in details.

Part 2 contains blocks for column and row selections of the input matrices. Also, block for the Hermite transposition of the CS matrix is used as well as circuits for matrix multiplications.

Part 3 contains register with the measurement vector values, matrix obtained as product of the CS matrix and transposed CS matrix multiplication, circuits for matrix inversion and matrix multiplication, two registers for storing intermediate result of the multiplication and vector obtained after the optimization problem solving.

### IV. HARDWARE ARCHITECTURE OF THE THRESHOLD-BASED ALGORITHM FOR CS RECONSTRUCTION

The architecture for the algorithm implementation is in detail described in the sequel. The scheme is consisted of several blocks. *Block 1* is used for random selection of $M$ out of $N$ samples of the signal $x$ at the input of the random generator. The randomly selected samples are stored in the vector $v$ at the output of the generator. The positions of the randomized signal are stored in the vector $P_v$. The available set of samples $v$ is brought to the input of the circuit for Fourier transform calculation (*Block 2* on the Fig.1). The FFT vector obtained at the output of the FFT block is called the initial FFT vector and it is fed at the input of the comparator block (*Block 3*, Fig.1). Threshold, calculated in the *Block 4*, is also fed at the input of the *Block 3*. Input samples of the initial FFT are compared with the threshold. Fig.1b shows initial FFT vector and threshold for the signal consisted of 14 components, defined according to the (2). The results are obtained using MATLAB, in order to better illustrate the functionality of the proposed block scheme.

If value of the FFT sample is above the defined threshold, output of the comparator circuit gives "1", otherwise, the output is "0". These values will be used for the column selection of the DFT matrix in part 2 of the scheme. The formation of the CS matrix is shown in Fig. 2. The comparator outputs are brought to the input of the *Block 5* (Fig.2), together with the DFT matrix $A$. Matrix $A$ is column sub-sampled: columns whose indexes correspond to the indexes of "1" in $Cr$ are selected, while the others are discarded. After the column subsampling, row subsampling is performed. The rows that correspond to the $M$ indexes of the randomly permuted positions $P_v$ (Fig. 1) are selected. At the output of the row selection block, CS matrix $A_{CS}$ is obtained. Row dimension $M$ of the matrix corresponds to the number of the random measurements while the column dimension is equal to the number of FFT samples that are above the threshold. After the matrix transpose (*Block 7*) and multiplication with the original matrix $A_{CS}$, the $A_P$ matrix of $K\times K$ dimension is obtained.

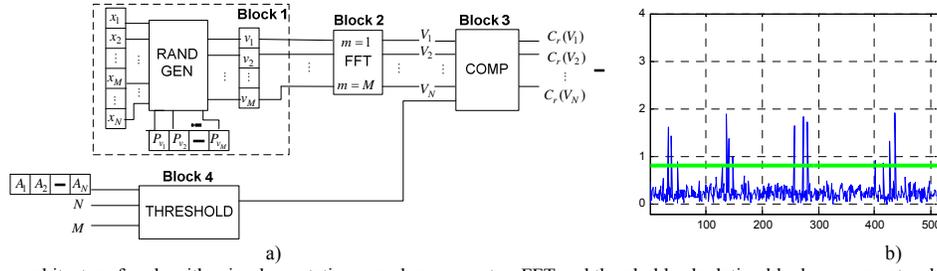

Figure 1: a) Part of the architecture for algorithm implementation - random generator, FFT and threshold calculation blocks, comparator; b) An illustration of the initial DFT and threshold

Note that the matrix transpose in the FPGA can be done by transferring column-wise data from the memory to the chip. After that, the column data are realigned into adjacent addresses, and stored back to the memory [21].The $X_P$ vector is obtained by multiplying transposed $A_{CS}$ matrix with the measurement vector $v$. Multiplication of the vector $X_P$ with the inverse form of the matrix $A_P$ produces the vector $X_{TP}$, containing exact DFT amplitudes of the signal components. Matrix inversion can be calculated by using recursive least square algorithm based on QR decomposition [22]. In order to obtain full, $N \times 1$ DFT of the reconstructed signal, the samples have to be arranged to the corresponding frequency positions. In that sense, block for the spectral positioning is used (Fig. 3). As input, this block uses values of the vector $X_{TP}$ and matches these values to the corresponding frequencies. The rest ($N$-$K$) vector positions are filled with zeros. The resulting $X_{N \times 1}$ vector is DFT of reconstructed signal. Time domain reconstructed signal can be obtained by applying IDFT. For the illustration of the results, the original and reconstructed DFT is shown in Fig. 3b (the results are depicted in MATLAB).

Block for the threshold calculation is shown in the Fig. 4. At the input of the adders the signal length $N$, number of available samples $M$, amplitudes of the signal components and probability $P$ is fed. As a result, the threshold is obtained.

### A. The logarithmic and power function calculation

Having in mind that in FPGA there are no standard components for the logarithmic and power functions, these functions can be calculated using CORDIC algorithm, Look-up-Tables (LUT), or polynomial approximations [13], [18]. The LUT approach for the logarithmic function calculation will be described in the sequel, as this approach provides good precision and high speed. Logarithm with the base 10 can be written as:

$$\log_{10} x = \frac{\log_2 x}{\log_2 10}. \qquad (8)$$

If $x$ is observed as floating point number, with $x_m$ mantissa, and $x_e$ exponent, i.e., $x=x_m 2^{x_e}$, previous relation becomes:

$$\log_{10} x = \frac{\log_2(x_m 2^{x_e})}{\log_2 10} = \frac{\log_2(x_m) + x_e}{\log_2 10}. \qquad (9)$$

The $log_2(x_m)$ can be calculated by using the LUTs, created as: $LUT(x_m)=round(2^{15}log_2(x_m))$. Therefore, relation becomes:

$$\log_2(x_m 2^{x_e}) = x_e + \log_2(x_m) = x_e + LUT(x_m). \qquad (10)$$

Let us discuss square root implementation. There is no component for square root calculation in FPGA, and several algorithms for its calculation exist. They can be divided in two groups: first - Rough estimation and Newton-Raphson method and second - digit-by-digit method (restoring and non-restoring algorithms). Here, the non-restoring algorithm will be described, as it can be implemented with fewer hardware resource compared to the restoring algorithm. If $B$ is 32-bit unsigned number whose square root is founding, $R$ is 17-bit reminder and $W$ is the 17-bit result, then the pseudo code can be as follows [19]:

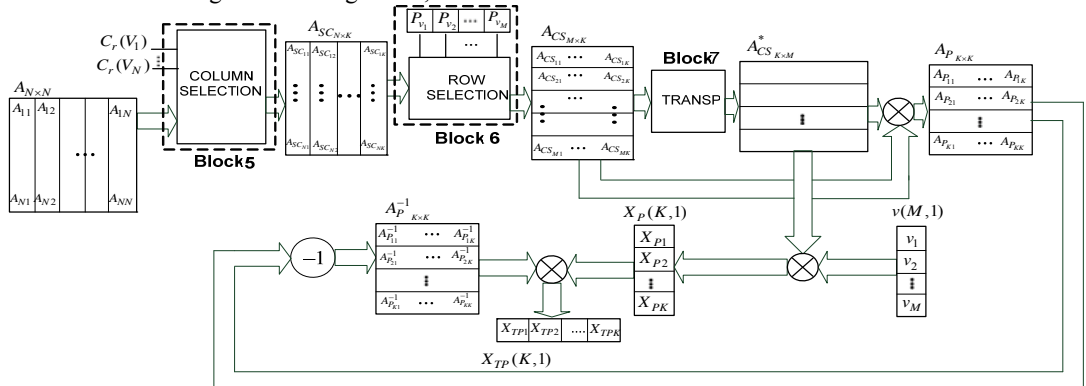

Figure 2: Parts 2 and 3 of the block scheme for algorithm FPGA implementation

```
for i=15:-1:0
    w_16=0, r_16=0
        if r_{i+1}>=0 then r_i=r_{i+1}B_{2i+1}B_{2i}-w_{i+1}01
                      else r_i=r_{i+1}B_{2i+1}B_{2i}+w_{i+1}11
        end
        if r_i>=0 then w_i=w_{i+1}1
                  else w_i=w_{i+1}0
        end
end
if r_0<0 then r_0= r_0+ w_0 1
```

Vector $w_i$ has (16-i) bits: $w_i=W(15:i)$, $w_0= W(15:0)$ and $r_0= R(16:0)$. The expressions $(r_{i+1}B_{2i+1}B_{2i})$ and $(w_{i+1}1)$ denotes:

$$r_{i+1}B_{2i+1}B_{2i} \to r_{i+1} \times 2^2 + B_{2i+1} \times 2^1 + B_{2i} \times 2^0$$
$$w_{i+1}1 \to w_{i+1} \times 2^1 + 2^0 \quad (11)$$

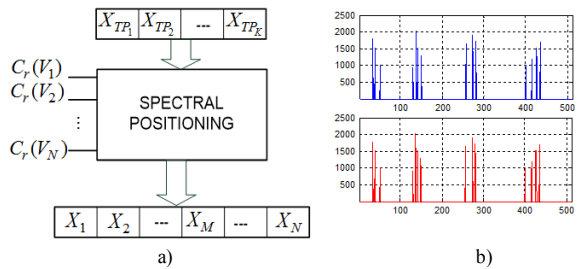

Figure 3: a) Block for the spectral positioning of the reconstructed signal amplitudes; b) Illustration of original (blue) and reconstructed (red) DFT

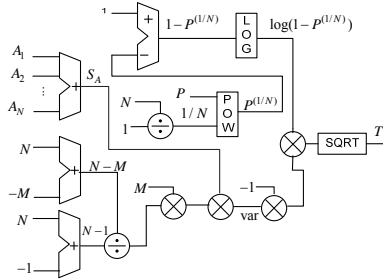

Figure 4: Architecture for the threshold calculation

## V. CONCLUSION

The architecture for hardware implementation of the single iteration algorithm allowing fast and efficient CS reconstruction is proposed. This system is suitable for FPGA implementation required for real-time solutions. The hardware scheme is divided into three parts, while the building blocks for each part are analyzed and described in terms of the functionality. Also, for the blocks that are not based on the standard FPGA components, a suitable FPGA implementation is described as well.

## REFERENCES

[1] R. Baraniuk, "Compressive Sensing", Lecture Notes in IEEE Signal Processing Magazine, vol. 24, July 2007.

[2] I. Orovic, S. Stankovic, T. Thayaparan, "Time-Frequency Based Instantaneous Frequency Estimation of Sparse Signals from an Incomplete Set of Samples," IET Signal Processing, Special issue on Compressive Sensing and Robust Transforms, May, 2014.

[3] M. Elad, "Sparse and Redundant Representations: From Theory to Applications in Signal and Image Processing", Springer 2010.

[4] M. F. Duarte, Y. C. Eldar, "Structured Compressed Sensing: From Theory to Applications", IEEE Transactions on Signal Processing, Vol. 59, No. 9, September 2011.

[5] S Stanković, I Orović, M Amin, "L-statistics based modification of reconstruction algorithms for compressive sensing in the presence of impulse noise", Signal Processing, vol. 93, no. 11, pp. 2927-2931, 2013.

[6] S. Stankovic, LJ. Stankovic, I. Orovic, "Relationship between the Robust Statistics Theory and Sparse Compressive Sensed Signals Reconstruction," IET Signal Processing, Special issue on Compressive Sensing and Robust Transforms, May, 2014.

[7] M. F. Duarte, R. G. Baraniuk, "Recovery of frequency-sparse signals from compressive measurements," 48th Annual Allerton Conference on Communication, Control, and Computing (Allerton), 2010, pp.599-606, Sept. 29-Oct. 1 2010.

[8] E. Candes, J. Romberg, "l1-magic: Recovery of Sparse Signals via Convex Programming", October 2005.

[9] T. Blumensath, M. E. Davies, "Gradient Pursuits," IEEE Transactions on Signal Processing, vol.56, no.6, pp.2370-2382, June 2008.

[10] S. Stankovic, I. Orovic, LJ. Stankovic, "An Automated Signal Reconstruction Method based on Analysis of Compressive Sensed Signals in Noisy Environment," Signal Processing, vol. 104, pp. 43 - 50, 2014

[11] M. A. T. Figueiredo, R. D. Nowak, S. J. Wright, "Gradient Projection for Sparse Reconstruction: Application to Compressed Sensing and Other Inverse Problems," IEEE Journal of Selected Topics in Signal Processing, vol.1, no.4, pp.586-597, Dec. 2007

[12] S. Stankovic, I. Orovic, LJ. Stankovic, A. Draganic, "Single-Iteration Algorithm for Compressive Sensing Reconstruction," Telfor Journal, Vol. 6, No. 1, pp. 36-41, 2014

[13] P. K. Meher, J. Valls, Tso-Bing Juan, K. Sridharan, K. Maharatna, "50 Years of CORDIC: Algorithms, Architectures, and Applications," IEEE Transactions on Circuits and Systems I: Regular Papers, vol.56, no.9, pp.1893-1907, Sept. 2009.

[14] S. Stankovic, I. Djurovic, V. Vukovic, "System architecture for space-frequency image analysis," Electronics Letters, Vol. 34, No.23, pp.2224-2245, 1998.

[15] I. Orovic, S. Stankovic, B. Jokanovic, "A Suitable Hardware Realization for the Cohen Class Distributions," IEEE Transactions on Circuits and Systems II, no.99, pp.1-5, Avg., 2013.

[16] L. Bai, P. Maechler, M. Muehlberghuber, H. Kaeslin, "High-speed compressed sensing reconstruction on FPGA using OMP and AMP," 19th IEEE International Conference on Electronics, Circuits and Systems (ICECS), 2012, pp.53,56, 9-12 Dec. 2012.

[17] N. Zaric, N. Lekic, S. Stankovic, "An Implementation of the L-estimate Distributions for Analysis of Signals in Heavy-Tailed Noise," IEEE Trans. on Circuits and Systems II, Vol 58, No. 7, pp. 427-432, 2011.

[18] J.-A. Pineiro, M.D. Ercegovac, J.D. Bruguera, "Algorithm and architecture for logarithm, exponential, and powering computation," IEEE Transactions on Computers, vol.53, no.9, pp.1085,1096, Sept. 2004, doi: 10.1109/TC.2004.53.

[19] Y. Li, W. Chu, "Implementation of single precision floating point square root on FPGAs," Proceedings of the 5th Annual IEEE Symposium on Field-Programmable Custom Computing Machines, 1997., pp.226,232, 16-18 Apr 1997.

[20] E. E. Swartzlander, H. H.M. Saleh, "FFT Implementation with Fused Floating-Point Operations" IEEE Trans. on Computers, vol. 61, no. 2, 2012, pp. 284-288.

[21] Y. Chi-Li Yu, C. Chakrabarti, "Transpose-free SAR imaging on FPGA platform," IEEE International Symposium on Circuits and Systems (ISCAS), 2012, pp.762, 765, 20-23 May 2012.

[22] M. Karkooti, J. R. Cavallaro, C. Dick, "FPGA Implementation of Matrix Inversion Using QRD-RLS Algorithm," Conference Record of the Thirty-Ninth Asilomar Conference on Signals, Systems and Computers, 2005., pp.1625,1629, October 28 - November 1, 2005.